\documentclass[useAMS,usenatbib]{mn2e}
\usepackage{mn2ejour}
\usepackage{amssymb}
\usepackage{color}
\usepackage{gensymb}
\usepackage{graphicx}

\title[Water delivery]{Water delivery to the TRAPPIST-1 planets}

\author[Z. Dencs \& Zs. Reg\'aly]{Z. Dencs$^{1,2}$\thanks{E-mail: dencs.zoltan@csfk.mta.hu} and Zs. Reg\'aly$^{1}$\\
$^1$Konkoly Observatory, Research Centre for Astronomy and Earth Sciences, Hungarian Academy of Sciences,\\\,\,\,\,1121, Budapest, Konkoly Thege Mikl\'os \'ut 15-17, Hungary\\
$^2$Department of Astronomy, E\"otv\"os Lor\'and University, Budapest, Hungary, H-1117}

\begin{document}

\pagerange{\pageref{firstpage}--\pageref{lastpage}} \pubyear{2018}

\maketitle
\label{firstpage}

\begin{abstract}
Three of the seven rocky planets (e, f, and g) in TRAPPIST-1 system orbit in the habitable zone of the host star. Therefore, water can be in liquid state at their surface being essential for life. Recent studies suggest that these planets formed beyond the snow line in a water-rich region. The initial water reservoir can be lost during the planet formation due to the stellar activity of the infant low-mass star. However, a potential subsequent water delivery event, like the late heavy bombardment (LHB) in the Solar System, can replenish planetary water reservoirs. To study this water delivery process, we set up a simple model in which an additional 5\,M$_{\oplus}$--50\,M$_{\oplus}$ planet is embedded in a water-rich asteroid belt beyond the snow line of TRAPPIST-1. Asteroids perturbed out from the chaotic zone of the putative planet can enter into the inner system and accreted by the known planets. Our main finding is that the larger is the orbital distance of planet, the higher is the amount of water delivered to the planet by an LHB-like event. 
\end{abstract}
 

\hyphenation{REBOUND}
\hyphenation{TRAPPIST}
\hyphenation{HIPERION}

\section{Introduction}

TRAPPIST-1 is an M8 type 0.0802\,M$_{\odot}$ ultra-cool red dwarf star at distance of 12.1\,pc from the Solar System harbours seven known Earth-sized rocky planets \citep{Gillonetal2017}. By assuming that life requires liquid water, search for extraterrestrial life is focused on Earth-like planets that can have liquid water on their surface. Based on the stellar luminosity ($5\times10^{-4}$ L$_{\odot}$) of TRAPPIST-1 the liquid water habitable zone extends from 0.024\,au to 0.049\,au in the TRAPPIST-1 system \citep{Shieldsetal2016}. Within that region three terrestrial planets, TRAPPIST-1e, f, and g were discovered \citep{Gillonetal2017}. 

Recent theoretical models assume that TRAPPIST-1 planets formed ex situ, and they collected their mass during an inward migration event. These migrating models predict the formation of resonant chains, which is indeed observed identified in TRAPPIST-1 system \citep{Izidoroetal2017,Lugeretal2017,Unterbornetal2018}. 

In protoplanetary discs volatiles, such as H$_2$O, NH$_3$, and CH$_4$, condense beyond the snow line on silicate dust grains. TRAPPIST-1 planets forming beyond the snow line can collect significant amount of volatiles during their inward migration. However, according to \cite{Bolmontetal2017} a large amount of water can be lost during planet formation process due to the strong XUV radiation of a pre-main sequence low mass star. The early phase stellar activity can cause greenhouse effect on the planets, which may lead to the photodissociation of water molecules in the atmosphere and escaping of hydrogen. As a result, the primordial atmosphere of TRAPPIST-1 planets can be completely eroded over a few billion years due to the stellar activity \citep{Bourrieretal2017}. Note that the lack of hydrogen lines in the atmosphere of TRAPPIST-1 planets is confirmed recently by \cite{deWitetal2018}. As the existence of the atmosphere and the atmospheric water are confirmed by transmission spectra \citep{deWitetal2018}, a process explaining the formation of the secondary atmospheres and the origin of water is necessary.

An early Kuiper belt-like asteroid ring beyond the orbit of the outermost planet\,h in the TRAPPIST-1 system could have been survived the violent planet formation event \citep{Hahn2003}. This asteroid ring can mainly be consisted of volatile-rich bodies: volatile-to-solid ratio is estimated to be 50\,percent \citep{Schwarzetal2017}. An external planet perturbing the orbits of these bodies, may result in a scattering event whereby a significant amount of asteroids can enter into the inner system. A similar event is assumed to be occurred in our Solar System, known as the late heavy bombardment (LHB, \citealp{Gomesetal2005}). The concept of orbital instabilities of asteroids caused by a giant planet leading to an LHB-like event seems to be a common evolutionary stage for many planetary systems \citep{Thommesetal2008,Raymondetal2010,Raymondetal2011}. 

According to recent theoretical studies this inward scattering event results in the reproduction of sub-mm sized dust originated from an outer asteroid belt around main sequence stars \citep{Bonsoretal2012,Bonsoretal2014}. Due to the short life-time of mm-sized dust this reproduction process is still ongoing, which can be observed, e.g., in the case of $\kappa$\,CrB \citep{Bonsoretal2013}.

As in the case of extrasolar planetary systems the bombardment of water-rich planetesimals and planetary embryos could play an important role in the water delivery to the planets of the inner Solar System, which is occurred after the formation of Earth has been completed \citep{Morbidellietal2000,Raymondetal2007}. Recent models of   \citet{Bottkeetal2012} suggest that the bulk of the water-rich asteroids collided into Earth originates from the inner edge of the main asteroid belt. Other studies suggest that asteroids may come from the outer regions of the main asteroid belt, or even beyond it \citep{Walshetal2011,RaymondIzidoro2017,OBrienetal2018}.

There are some hints of the presence of an additional planet in the TRAPPIST-1 system: a large amplitude wobbling of the host star (1.9\,mas, \citealp{Bossetal2017}), and a significant oscillation are observed in the stellar radial velocity (150\,m/s, \citealp{Tanneretal2012}). According to \cite{Bossetal2017} the observed amplitude of stellar wobbling can be explained by less than 4.6\,Jupiter mass (M$_{\mathrm{Jup}}$) planet with one year orbital period, or less than 1.6\,M$_{\mathrm{Jup}}$ planet with five years period. However, the formation of these massive planets can hardly be explained by assuming a canonical protoplanetary disc mass. Note that by taking an analogy to Solar System's LHB, a few ten Earth mass (M$_{\oplus}$) planet can be considered as a giant planet in TRAPPIST-1 system (like Saturn in our Solar system).

With regards to the stellar radial velocity, we found that known planets cause only $\sim$35\,m/s velocity amplitude being approximately one fifth that of observed. A putative planet with 50\,M$_{\oplus}$ (largest mass assumed in our hypothetical system) orbiting at 0.117\,au (the tightest stable configuration, see details in Section\,2) can cause only $\sim$60\,m/s radial velocity amplitude, which is below the measured amplitude value. However, stellar activity can also affect radial velocity measurements. According to  \citet{AndersenKorhonen2015} the detection of a few M$_{\oplus}$ planet by radial velocity measurements requires more than a hundred observations (unavailable currently) due to the high activity level of stellar atmosphere of TRAPPIST-1 \citep{Vidaetal2017}.

In a recent analysis, \citet{Grimmetal2018} used 107 orbits of planet\,b (orbital period: 1.51\,days) and 7 orbits of planet\,h (18.77\,days period) for determining transit timing variations (TTV) signals. This means that only 4 and 1 orbits are available to detect by TTV signals our hypothetical planet that orbits at the tightest and widest configuration, respectively. Therefore, the presence of an additional planet orbiting in TRAPPIST-1 system may not be ruled out by the available TTV measurements.

The aim of this study is to investigate how does an additional planet in TRAPPIST-1 system, with a mass and orbital distance not contradicting to the available observations, excite an LHB-like event, in which water is transported to the known planets. 

Our investigation is based on numerical simulations of planet--asteroid gravitational interactions by means of N-body integrations. Section\,2 deals with a limited stability analysis of the extended planetary system. Section\,3 presents the N-body simulations for determining the asteroid impact flux on known TRAPPIST-1 planets. In Section\,4, we discuss the results of the planet--asteroid interaction simulations, and present the estimated amount of water that can be transported via asteroid accretion. In Section\,5, we present conclusions based on the results of our simulations.

\section{Numerical models}

\begin{figure}
	\begin{center}
	    \includegraphics[width=1\columnwidth]{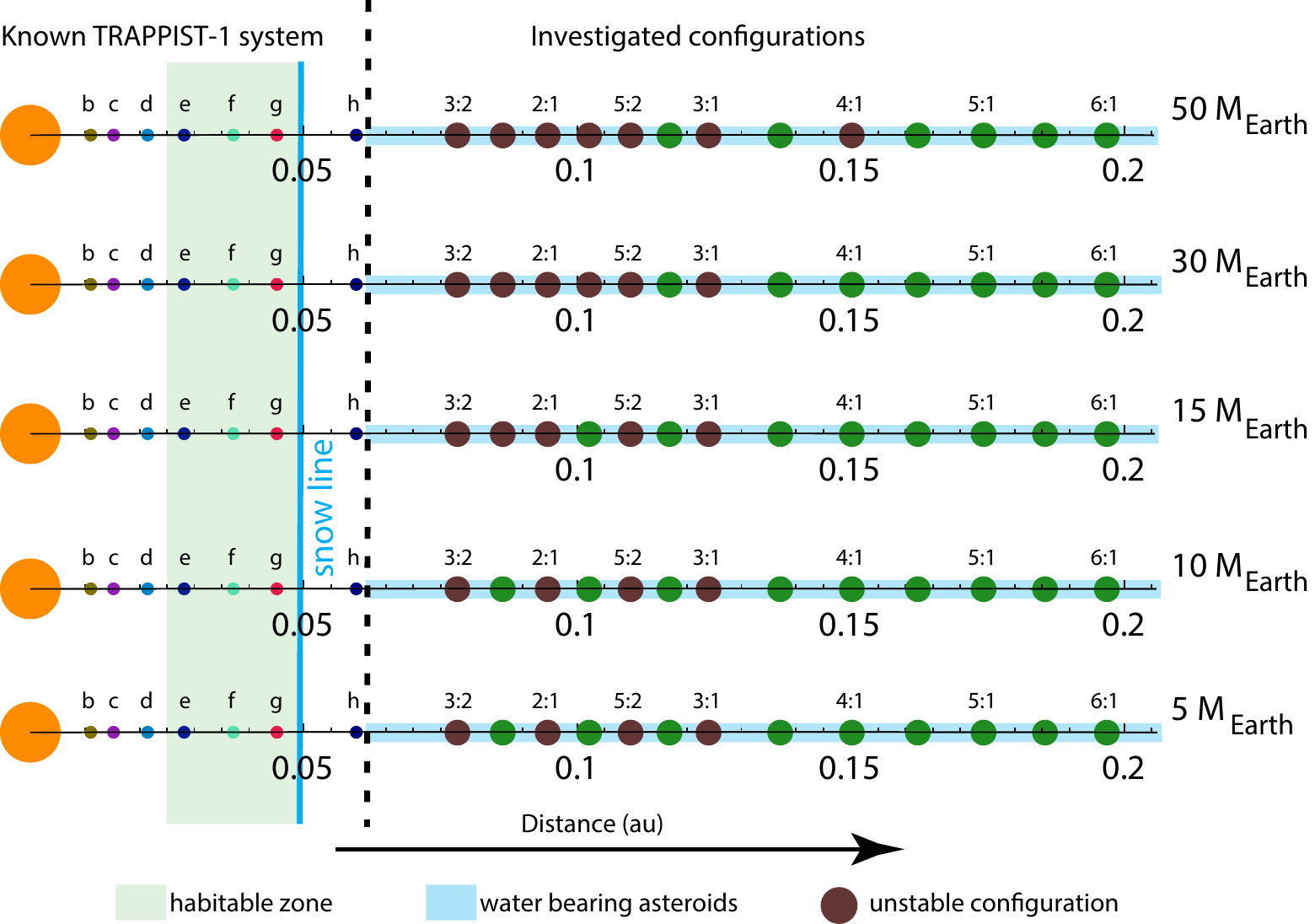}
	\end{center}
	\caption{Architecture of our hypothetical TRAPPIST-1 system assuming an additional 5, 10, 15, 30, and 50\,M$_{\oplus}$ putative planet\,i. Rows correspond to different mass of planet\,i. The blue and the green regions indicate the water-rich asteroid belt and the circum-stellar habitable zone, respectively. Green and brown circles indicate the stable and unstable systems, respectively.}
	\label{fig:configurations}
\end{figure}

We assume that asteroids colliding to TRAPPIST-1 known planets are originated from a water-rich belt, which is  perturbated by a putative several Earth mass planet (planet\,i) embedded in that asteroid belt. In order to investigate the asteroid impact flux on the known TRAPPIST-1 planets, the number of the asteroids accreted by a given planet ($N_{\mathrm{acc}}$) is measured in N-body simulations.

We investigated models in which planet\,i has a mass of $m_{\mathrm{pl, i}}=5$, 10, 15, 30, and 50\,M$_{\oplus}$. The semi-major axis of planet\,i ($a_{\mathrm{pl, i}}$) is set to be in the range of $0.078\,\mathrm{au} \leq a_{\mathrm{pl, i}} \leq 0.258\,\mathrm{au}$: in seven cases planet\,i is in mean motion resonance (MMR) with planet\,h (3:2, 2:1, 5:2, 3:1, 4:1, 5:1, 6:1); in six cases the orbital distance of planet\,i are between these resonances. The architecture of investigated models are shown in Fig.\,\ref{fig:configurations}. 

\subsection{Orbital stability analysis}

As a first step, we investigated the orbital stability of the model of TRAPPIST-1 system with the known planets. We used an open-source software package REBOUND \citep{ReinSpiegel2015,ReinTamayo2016,Tamayoetal2017}. We used IAS15 integrator of REBOUND with the planetary and the orbital parameters derived from the measured TTV light curve analysis \citep{Wangetal2017}. IAS15 applies non-symplectic 15$^{\mathrm{th}}$ order integrator and adaptive time-step. The energy error of the system normalised with the initial energy is in the order of $10^{-12}$. The simulations ran for two million years, $4.8\times10^8$ orbits of planet\,b, which corresponds to $3.85\times10^7$ orbits of planet\,h. We found that, the modelled system is stable as well as the eccentricity and the inclination of planet\,b--h stay in the limits given by the observational uncertainties of \citet{Wangetal2017}. 

As a next step in the stability analysis, planet\,i added to the system. The strength of the perturber's perturbation strongly depends on the mass and the distance of the perturber. We defined two types of initial orbital parameters and masses for the seven known planets: 1) the values of the known planet masses and semi-major axes are  based on \cite{Wangetal2017}, while the eccentricities and inclinations are zero (see parameters in Table\,\ref{tab:parameters}); 2) we applied initial eccentric models of \cite{Quarlesetal2017} shown to be the most stable configuration. As the conjunction of more than two planets results an unstable configuration, the initial position of planets (true anomaly) are set such that the conjunction of any planets is avoided. Laplace resonance provide to avoid of these conjunctions. Note that three body Laplace resonances indeed occur in the TRAPPIST-1 system, i.e. any two of the three nearby planets are in first-order MMR \citep{Papaloizouetal2018}.

\begin{table}
\begin{center}
\caption{Parameters for TRAPPIST-1 known planets according to \citet{Wangetal2017}.}
\label{tab:parameters}
\begin{tabular}{cccc}
\hline
 planet &
 semi-major axis (au) &
 mass (M$_{\oplus}$) &
 radius (R$_{\oplus}$)
 \\
\hline
 b & 0.01111 & 0.79 & 1.086\\
 c & 0.01522 & 1.63 & 1.056\\
 d & 0.02145 & 0.33 & 0.772\\
 e & 0.02818 & 0.24 & 0.918\\
 f & 0.0371 & 0.36 & 1.045\\
 g & 0.0451 & 0.566 & 1.127\\
 h & 0.0596 & 0.086 & 0.715\\
\end{tabular}
\end{center}
\end{table}

Stable and unstable models are shown in Fig.\,\ref{fig:configurations} with green and brown circles representing plant\,i, respectively. Stability tests show that systems are stable if planet\,i orbits at larger distances than 0.17\,au (5:1 MMR with TRAPPIST-1h). Systems can also be stable at smaller orbital distance of planet\,i, however, it requires less than 15\,M$_{\oplus}$ planet\,i orbiting between MMRs. Our simulations show that the stability of the system is independent whether eccentric \citep{Quarlesetal2017} or circular initial configuration is used.

In stable configurations, the semi-major axes of TRAPPIST-1 planets show no significant variations: in the case of the strongest perturbation, the average variation of the planets' semi-major axes is about 0.13\%. However, the eccentricities of TRAPPIST-1 planets are more sensitive to the perturbations of planet\,i. In some stable configurations, the eccentricities of TRAPPIST-1d, e and f temporarily grow above the upper limit given by TTV analysis of \cite{Wangetal2017}, however stay below e=0.035. Note that these eccentricity peaks are short term variations, lasts about $10^5$ orbits for planet\,b corresponding to several thousand years.

In unstable configurations, the orbital elements of one or two planets are strongly perturbed causing complete disruption of the system. The life-time of a stable system shortens with increasing mass and decreasing orbital distance of planet\,i. However, systems where planet\,i orbits at 3:1 MMR have shorter life-time than systems in which planet\,i orbits at 2:1 MMR. 

With regards the observability of transit signal the orbital inclination of planet\,i can be a critical parameter if the angle between the ascending node and the line of sight is close to $\pm90\degree$. We found that the orbital inclinations of the TRAPPIST-1 planets do not grow above the values given by TTV analysis of \citet{Wangetal2017} in most of our models assuming $5\degree$ inclined orbit for planet\,i. However, systems with 30\,M$_{\oplus}$ or 50\,M$_{\oplus}$ planet\,i on $5\degree$ inclined orbit are found to be unstable if planet\,i orbits at 4:1 MMR or at 5:1 MMR, respectively.

\section{Results}

\subsection{Planet--asteroid interactions}

In order to measure the asteroid impact flux on the habitable planets, we ran N-body models in which the putative planet\,i is embedded in an asteroid belt. Parameters of known planets used for these simulations are listed in Table\,1.  The asteroid belt consists of half a million massless particles ($N_{\mathrm{tot}}=0.5\times10^6$). Since we used massless test-particle approximation for the asteroids planetesimal-driven migration of planets is ignored, which is thought to enhance inward scattering of outer planetesimals \citep{Bonsoretal2014,RaymondBonsor2014}. The initial radial distribution of asteroids is set such that their surface mass density is proportional to $R^{-1}$ taking into account the sub-mm observations of debris discs \citep{WilliamsCieza2011}. The asteroid belt spatial extension is fixed in all models assuming different orbital distance and mass for planet\,i. The inner and the outer edge of the asteroid belt are defined by the outer edge of planet\,h's chaotic zone (0.0613\,au) and the outer edge of the chaotic zone of the largest mass putative planet orbiting at 6:1 MMR with planet\,h (0.2564\,au), respectively. Here the size of the chaotic zone is calculated according to \cite{Wisdom1980}. 

As a result of the gravitational perturbation of planet\,i asteroids can be accreted by the perturber itself, scattered out from the system, or scattered into the inner planetary system. The asteroids entered to the inner system can be accreted or scattered out by the known planets. Asteroids are considered to be accreted (and removed from the system) if their distance from the centre of a given planet is less than the planetary radius. Asteroids are considered to be ejected from the system if the their distance from the centre of the system reaches up to 1\,au. In addition,  asteroids that approach the central star less than 0.01\,au are also removed from the system. 

For this investigation we used the GPU-based direct N-body code HIPERION\footnote{http://www.konkoly.hu/staff/regaly/research/hiperion.html} (HIgh PERformance Integrator fOr Nbody) applies a 6$^{\mathrm{th}}$ order Hermite integrator. Adaptive shared time-step method is applied using a second-order Aarseth scheme \citep{PressSpergel1988} with $\eta=0.025$. With this setting, the total energy of the modelled systems are conserved with a relative energy error of $10^{-9}$. The host star and 8 planets are interacting with each other and the massless asteroid particles: the N-body problem can be considered as half a million 9 plus 1 body problem. The length of the simulations are set to $3\times10^4$ orbits of planet\,h (corresponding to $3.6\times 10^5$ orbits of planet\,b). 

In majority of our models the inclination of the planetary orbits are set to zero, as well as the initial mean eccentricity ($<e>$) and inclination ($<i>$) of the asteroids. We call these models as dynamically cold models. In these models the accretion numbers are saturated by the end of the simulations, i.e., the LHB-like event  finished on a short time-scale. However, if we consider a dynamically hot model, where the asteroids initial eccentricities and inclinations are non-zero the LHB-like event takes longer. The dynamically hot asteroid belt simulations run about ten times longer than the cold models, see details in the next Section.

In order to investigate the length of the saturation time-scale of LHB-like event in a system assuming a 15\,M$_{\oplus}$ perturber planet\,i orbiting at 0.117\,au, we ran three additional simulations in which the asteroids initially orbit on eccentric and inclined orbits. We assume that the initial orbital elements of asteroids follow Rayleigh distribution according to \citep{IdaMakino1992}. In these models, the initial mean eccentricity and inclination of the asteroids' orbits are $<e>=10^{-4}$, $ 10^{-3}$, $10^{-2}$ and $<i>=0.5\times10^{-4}$, $0.5\times10^{-3}$, $0.5\times10^{-2}$. In the following we call these eccentric models dynamically hot models.

\subsection{Dynamically hot or cold asteroid belts}

\begin{figure}
	\begin{center}                                                         
        \includegraphics[width=1\columnwidth]{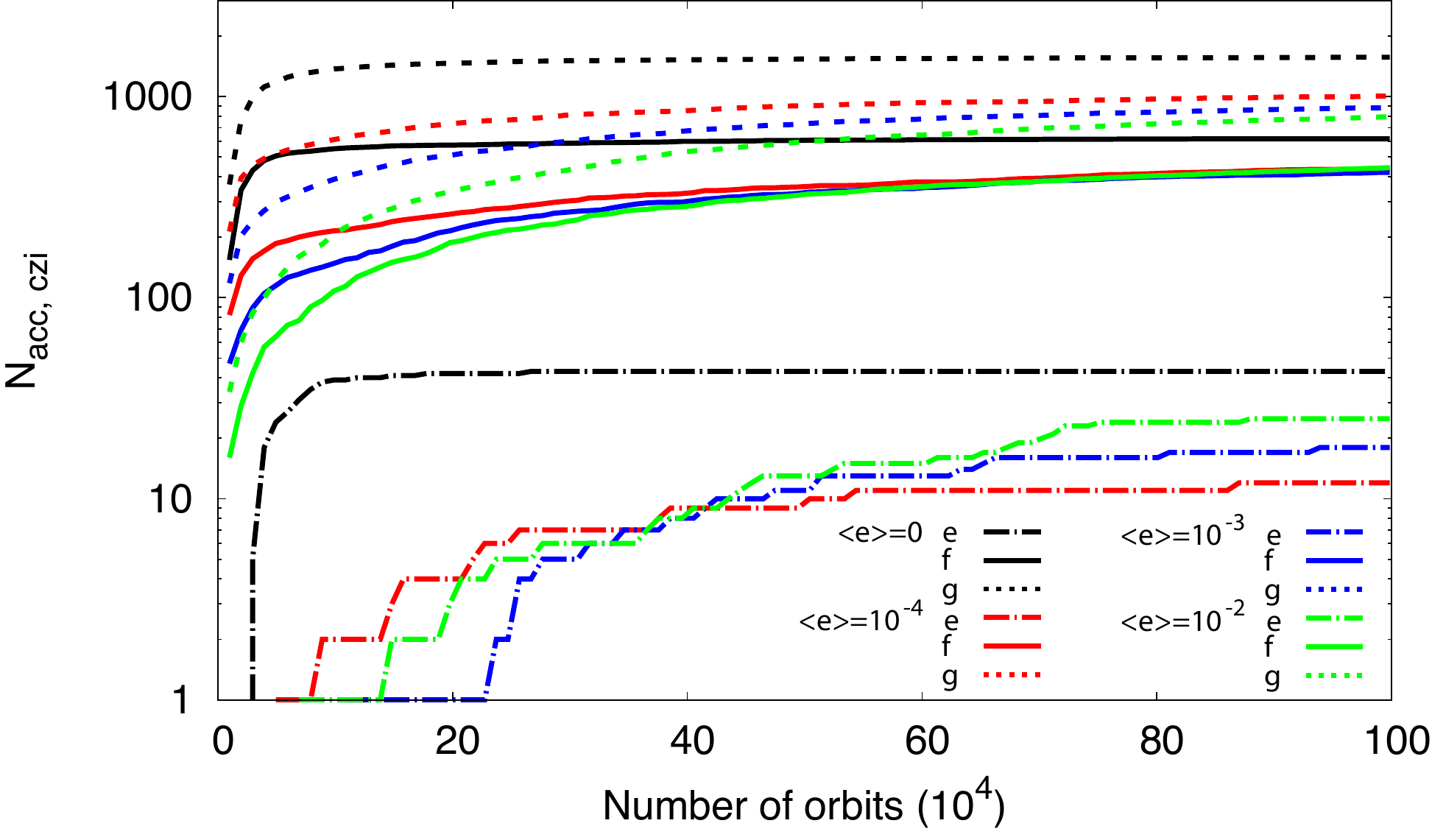}
	\end{center}
	\caption{ Number of accreted asteroids originated from the chaotic zone of planet\,i as a function of time for TRAPPIST-1e (dash-dotted lines), f (solid lines), and g (dotted lines) assuming a 15\,M$_{\oplus}$ planet\,i orbiting between 5:2 and 3:1 MMR at 0.117\,au. Time is measured by the number of orbits of planet\,b. Red, blue, and green colours correspond to models assuming $<e>=0$, $10^{-4}$, $10^{-3}$, and $10^{-2}$, respectively.}
	\label{fig:indiacc}
\end{figure}

\begin{figure}
	\begin{center}                                                         
        \includegraphics[width=1\columnwidth]{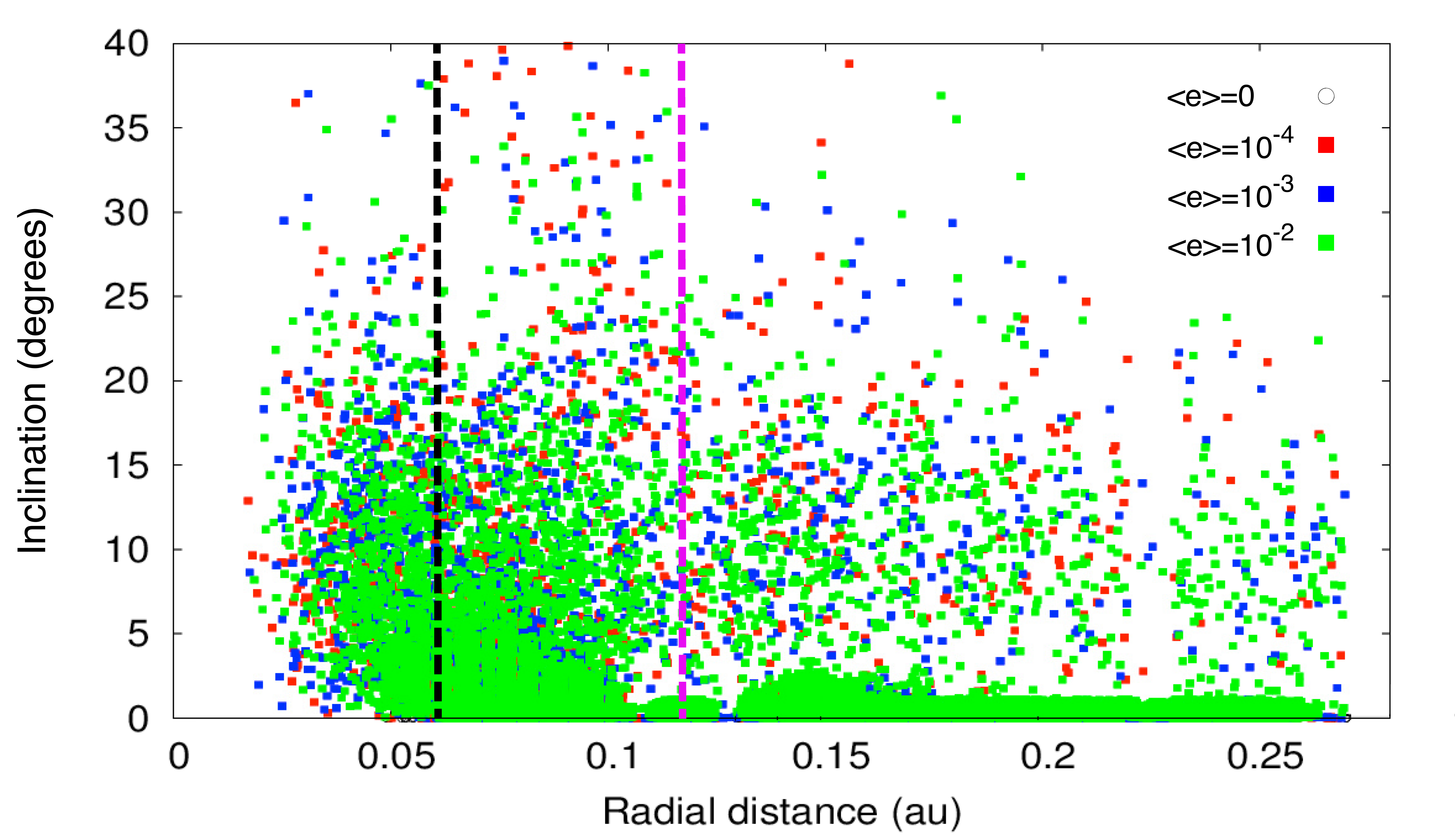}
    \end{center}
	\caption{Orbital inclination of the asteroids' orbits as a function of radial distance from TRAPPIST-1 after $3\times10^6$ orbits of planet\,b in the three dynamically hot models. Colours represent the same models as Fig\,\ref{fig:indiacc}. Black and purple dashed lines indicate the boundary between the inner and outer part of the system and the orbital distance of planet\,i, respectively.}
	\label{fig:inclination}
\end{figure}

Asteroids are scattered into the inner system from dynamically special regions of the asteroid belt, i.e. from chaotic zone of planet\,i, the 1:2 MMR with planet\,i, and from the inner edge of the belt, which is set next to the outer edge of planet\,h's chaotic zone. The time-scale of the gravitational perturbation of planet\,i is longer in the two latter regions than in the chaotic zone of planet\,i because they lie at larger distance from planet\,i. Therefore, asteroids originated from the chaotic zone of planet\,i enter into the inner system very fast. However, asteroids originated from the 1:2 MMR and belt edge are continuously pass into the inner system with a very slow rate. As a result, first a heavy bombardment phase occurs, which is followed by a low rate accretion phase.

Fig.\,\ref{fig:indiacc} shows the number of accreted asteroids originated only from planet\,i's chaotic zone ($N_{acc, czi}$) as a function of time for the habitable TRAPPIST-1 planets e, f, and g. As one can see, $N_{acc, czi}$ depends on the initial $<e>$ and $<i>$ of the asteroids and $N_{acc, czi}$ is about 2 times higher in cold model than that of in hot model. Moreover, $N_{acc, czi}$ only weekly depends on $<e>$, if $<e>$ greater than zero. We compared the sum of total number of accreted asteroids by TRAPPIST-1 known planets and the asteroids perturbed into the inner system in hot and cold models. We found that those numbers are the same (within several percent) in hot and cold models. Moreover, by extrapolating the depletion of asteroids from the inner system, we found that the time required to empty the inner system in hot models is approximately 50 times longer than that of in cold model.

The above finding can be explained by the following train of thought. The inclinations of the asteroids' orbits can only be excited by the perturbation of planet\,i, if the asteroids' inclinations are non-zero initially. As a result, inclination of asteroids which are perturbed into the inner system can be excited. Fig.\,\ref{fig:inclination} shows the inclination of the asteroids as a function of the radial distance after $3\times10^6$ orbits of planet\,b in the three dynamically hot models. One can see that the inclination of the asteroids' can reach surprisingly high values, i.e., about $40^\circ$ in the inner part of the system. The average inclination angle is found to be about $7.5^\circ$ in hot asteroids belt models independent of the initial $<e>$ and $<i>$. Thus, the probability of orbital crossing of asteroids and planets significantly decreases compared to the dynamically cold asteroids belt model.

This means that although the saturation of accretion number requires longer time in dynamically hot models, it saturates at the same level as cold model. Thus, we conclude that cold disk models are equivalent to hot models in terms of the total number of  accreted asteroids in the LHB phase. Therefore, we present our results only for cold models in the following.

\subsection{Effect of orbital elements of planet\,i} 

In the following, we investigate the asteroid impact fluxes to planet\,e, f, g and h only as TRAPPIST-1b, c and d accrete negligible amount of asteroids. Note that TRAPPIST-1b, c and d orbit outside the habitable zone, thus they are irrelevant in our investigation.

\begin{figure}
	\begin{center}                                                         
        \includegraphics[width=1\columnwidth]{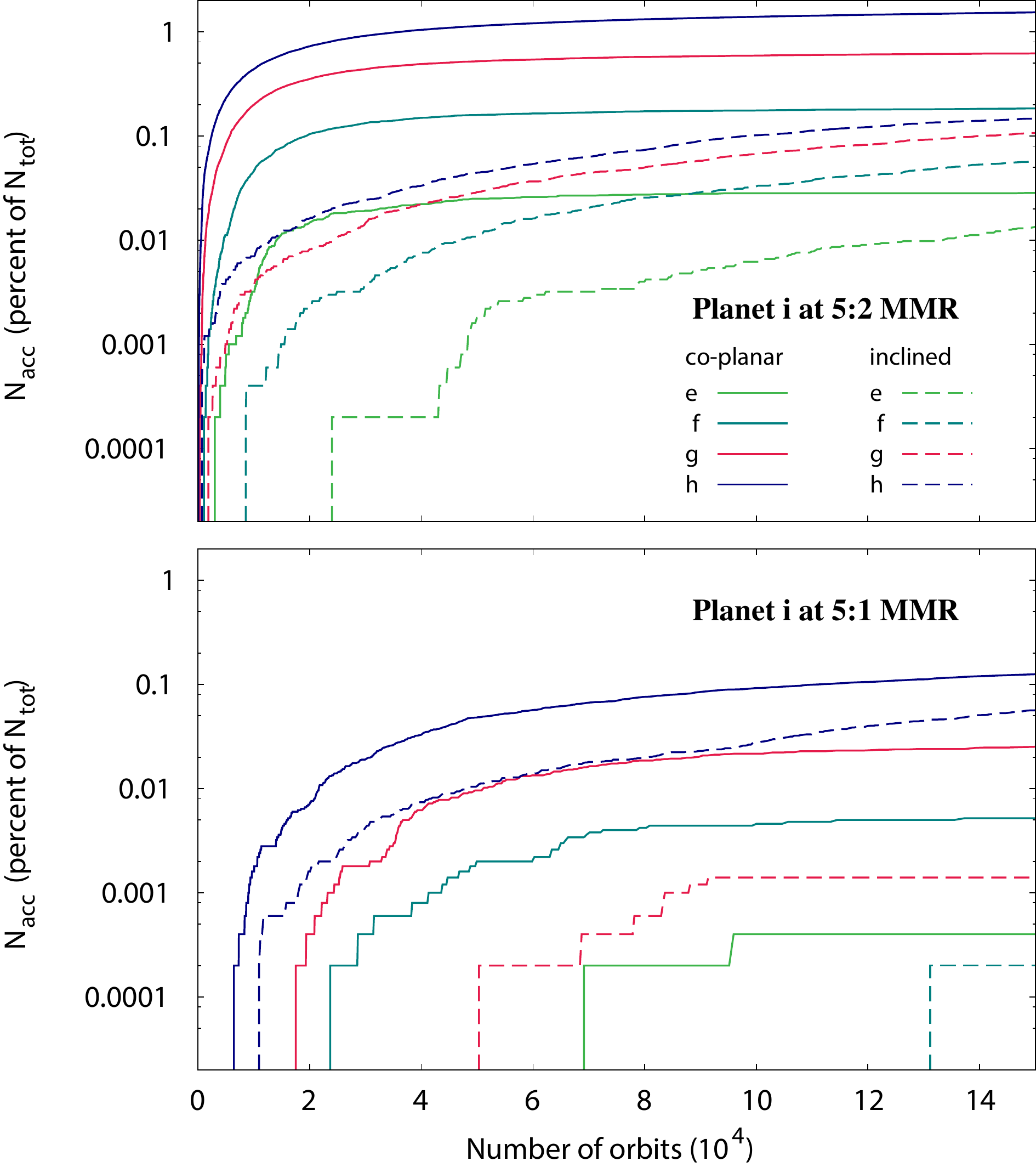}
    \end{center}
	\caption{$N_{\mathrm{acc}}$  as a function of time for TRAPPIST-1e, f, g and h assuming  15\,M$_{\oplus}$ planet\,i orbiting at 5:2 (upper panel) and 5:1 (lower panel) MMR with planet\,h. Solid and dashed lines correspond to planet\,i on a coplanar and inclined orbit, respectively. Time is measured by the number of orbits of planet\,b.}
	\label{fig:distanceacc}
\end{figure}

Fig.\,\ref{fig:distanceacc} shows the number of accreted asteroids on the habitable planets and planet\,h as a function of time in four different models: 1) 15\,M$_{\oplus}$ at 5:2 MMR (top panel, solid lines); 2) 15\,M$_{\oplus}$ at 5:1 MMR (bottom panel, solid lines); 3) 15\,M$_{\oplus}$ at 5:2 MMR on inclined (5$\degree$) orbit (top panel, dashed lines); 4) 15\,M$_{\oplus}$ at 5:1 MMR (bottom panel, dashed lines). Initially, the number of accreted asteroids rapidly grows, which is followed by a saturation phase lasting about a hundred thousand orbits of planet\,b. The time-scale of saturation decreases with the orbital distance of the accreting planet. By comparing the two panels of Fig.\,\ref{fig:distanceacc}, one can see that the saturation time-scale increases with the perturber's semi-major axis. The number of accreted asteroids ($N_\mathrm{acc}$) increases with the orbital distance of the accreting planet: the number of asteroids accreted by planet\,h is an order of magnitude larger than that of by planet\,f.

Comparing the coplanar to non-coplanar cases, it is also appreciable that $N_{\mathrm{acc}}$ is about one order of magnitude lower for models, where the perturber is on an inclined orbit with an inclination angle of $5\degree$ (see dashed lines in Fig.\,\ref{fig:distanceacc}). Note that planet\,i on an inclined orbit can excite the inclination of asteroids' orbits. As a result of the large mutual inclinations of the planets and the asteroids collisions are rare in non-coplanar models. For inclined models saturation of $N_\mathrm{acc}$ found to be prolonged beyond the simulation time.

We also investigated the dependence of $N_{\mathrm{acc}}$ on the initial orbital eccentricity of TRAPPIST-1 planets. In this investigation, the initial orbital eccentricities are set according to \citet{Quarlesetal2017}. We found that $N_{\mathrm{acc}}$ is not affected by the initial planetary eccentricities.

Furthermore, stable configurations with planet\,i on eccentric orbit ($e_{\mathrm{pl, i}}=0.1$ and 0.3) are also investigated. In these models $m_{\mathrm{pl, i}}=15$\,M$_{\oplus}$ planet\,i orbits at 4:1 MMR with planet\,h. We found that $N_{\mathrm{acc}}$ increases by a factor of about ten with the emerging eccentricity. Note, however, that these systems are found to be unstable. 

\subsection{Effect of the mass of planet\,i}

\begin{figure}
	\begin{center}                                                  
        \includegraphics[width=1\columnwidth]{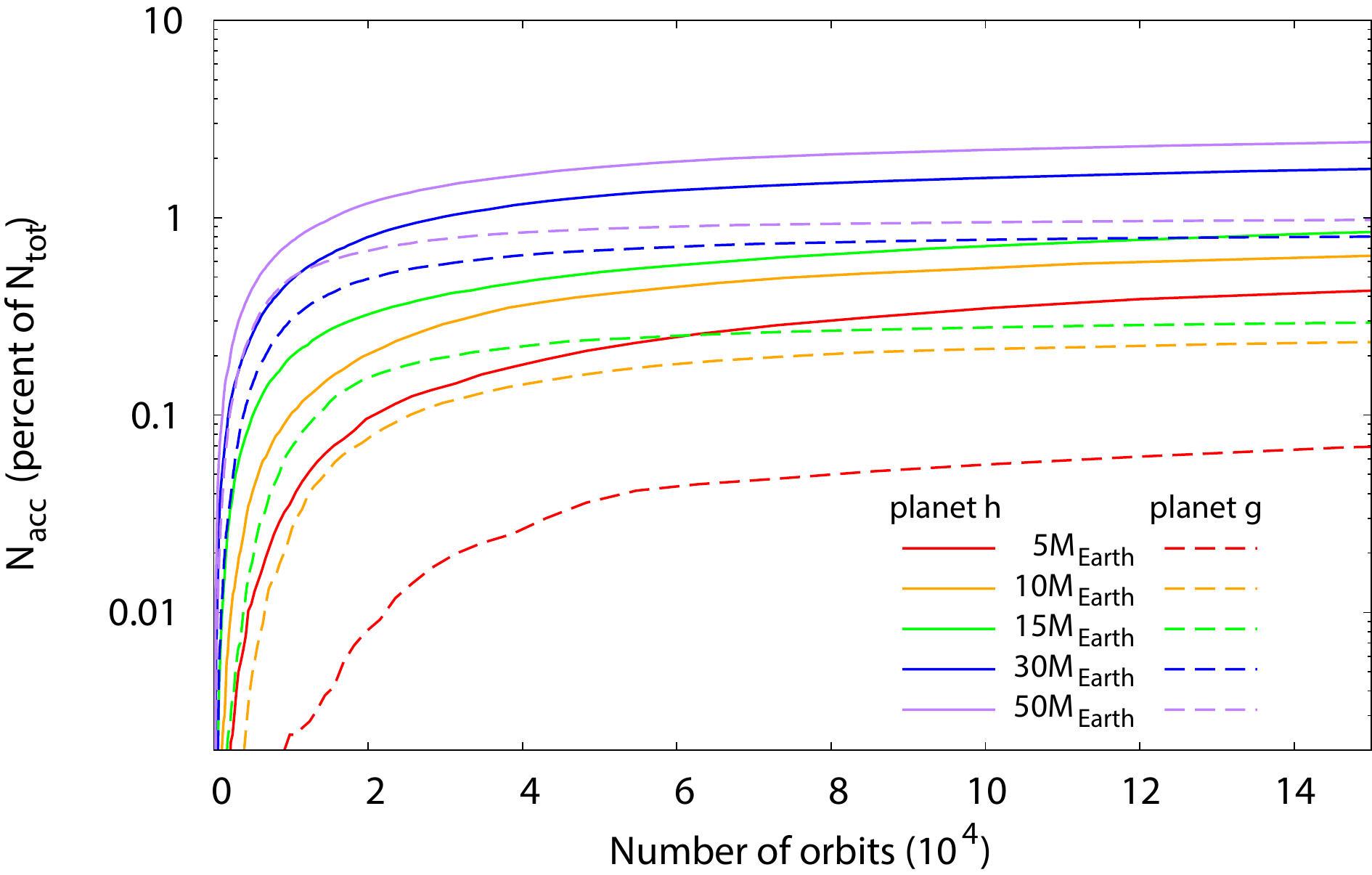}
    \end{center}
	\caption{$N_{\mathrm{acc}}$ as a function of time for TRAPPIST-1g (dashed lines) and h (solid lines) assuming 5--50\,M$_{\oplus}$ planet\,i orbiting at between 5:2 and 3:1 MMR at 0.117\,au. Time is measured by the number of orbits of planet\,b.}
	\label{fig:massacc}
\end{figure}

The effect of perturber's mass on the evolution of impact flux is presented in Fig.\,\ref{fig:massacc} showing $N_{\mathrm{acc}}$ on the two outermost known planets, planet\,g (dashed lines) and h (solid lines) versus time in the following models: 5--50\,M$_{\oplus}$ planet\,i orbiting at a distance of 0.117\,au (between 5:2 and 3:1 MMR) in coplanar configuration. It is obvious that the larger is the perturber's mass the stronger is the accretion. In all cases $N_{\mathrm{acc}}$ for planet\,h exceeds to that of planet\,g. It can also be seen that the saturation time-scale also depends on the perturber's mass: the larger is the perturber's mass the shorter is the saturation time-scale.

\begin{figure}
	\begin{center}                                                         
        \includegraphics[width=1\columnwidth]{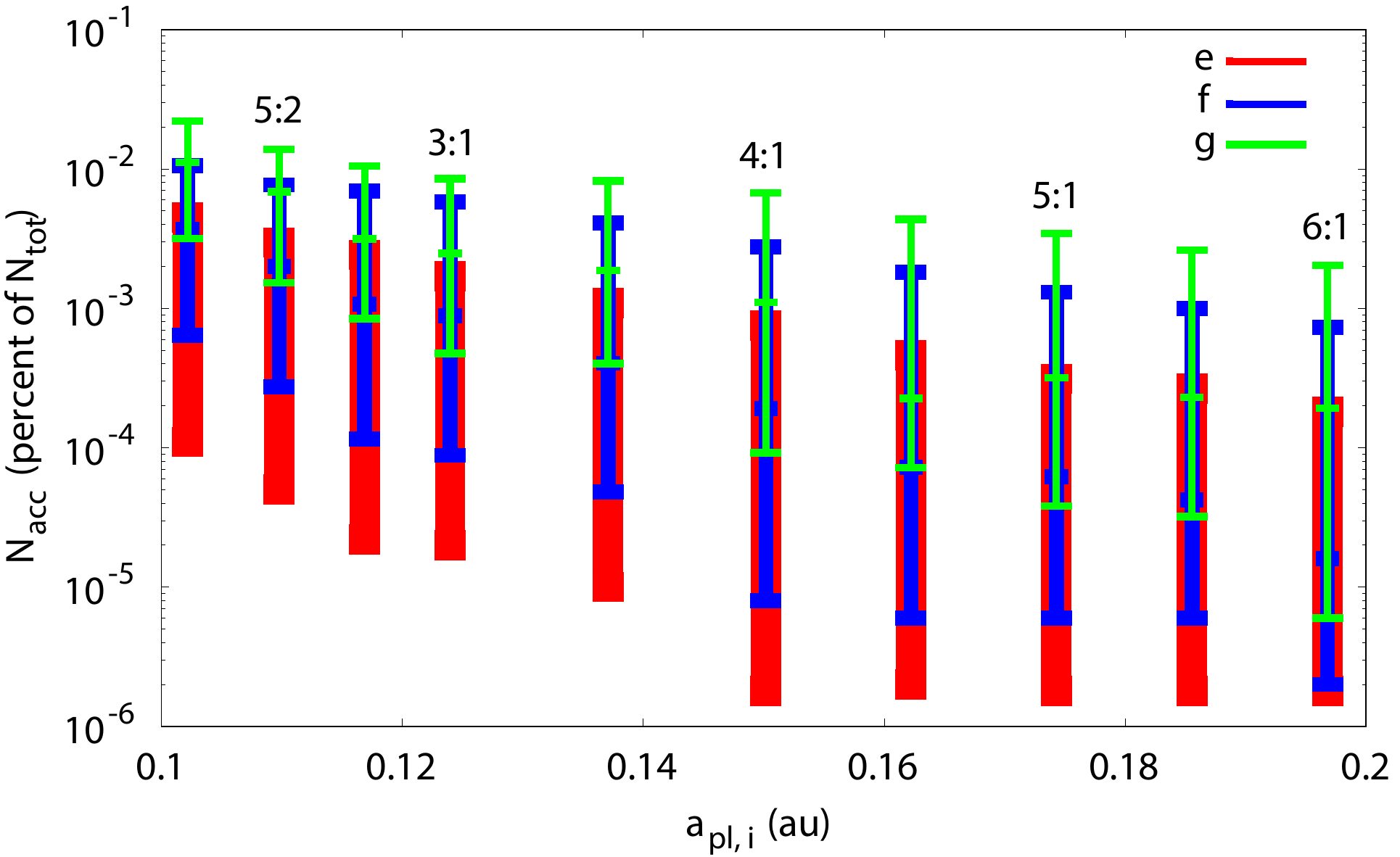}
	\end{center}
	\caption{Total number of the asteroids (as a percentage of the initial number of asteroids) accreted by the three habitable planets (e, f, and g) for different orbital distances of planet\,i on a coplanar orbit. Intervals correspond to models assuming different mass for planet\,i in the range of 5--50\,M$_{\oplus}$.}
	\label{fig:totalacc}
\end{figure}

Fig.\,\ref{fig:totalacc} shows $N_{\mathrm{acc}}$ on the habitable planets measured as  percentages of the total number of asteroids versus the semi-major axis of the perturber's orbit, $a_\mathrm{pl,i}$. The perturber's orbit in these models was again coplanar to the system. For each orbital distances 5 different perturber masses are investigated in the range of 5\,M$_{\oplus}\leq m_{\mathrm{pl, i}}\leq$ 50\,M$_{\oplus}$. Coloured ranges represent the scatter of $N_{\mathrm{acc}}$ which is due to the different perturber masses. Two major trends that were presented in Fig.\,\ref{fig:distanceacc} and \ref{fig:massacc} can also be identified in all cases. First, $N_{\mathrm{acc}}$ decreases with increasing $a_{\mathrm{pl, i}}$. Second, $N_{\mathrm{acc}}$ increases with the orbital distance of the accreting planets (e, f, and g). 

With respect to the asteroids are removed from the system, but not accreted by any planets, the following concerns can be drawn. The number of the asteroids accreted by the host star increases with $m_{\mathrm{pl, i}}$ and decreases with $a_{\mathrm{pl, i}}$. However, the number of asteroids ejected from the system increases with $a_{\mathrm{pl, i}}$. Note that the number of ejected asteroids is two order of magnitudes higher than that of accreted by TRAPPIST-1.

\section{Discussion}

\subsection{Overall intensity of the impact flux}

\begin{figure}
	\begin{center}                                                         
    	\includegraphics[width=1\columnwidth]{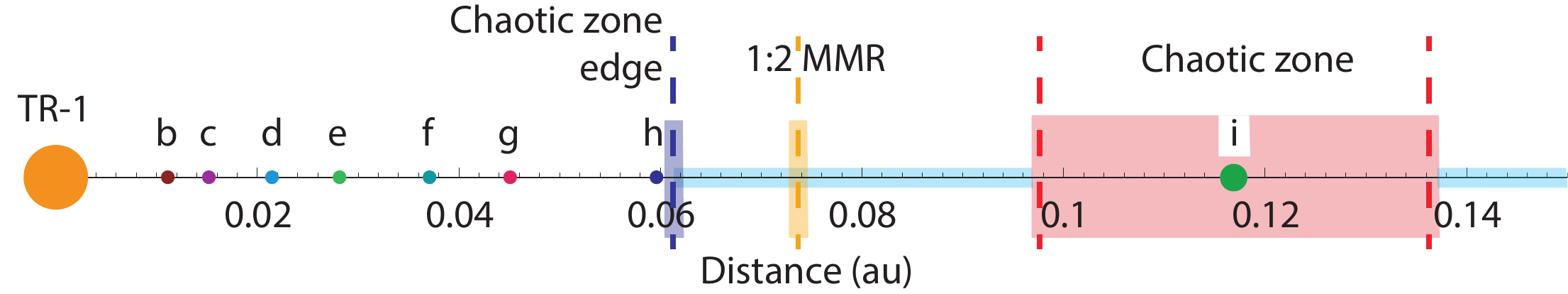}
    \end{center}
	\caption{Edge-on view of the model in which a 15\,M$_{\oplus}$ planet\,i orbits at 0.117\,au. Regions shaded with red (chaotic zone of planet\,i), orange (1:2 MMR with planet\,i) and blue (outer edge of planet\,h's chaotic zone) show the initial positions of asteroids accreted by the known planets. Light blue region shows the initial unperturbed asteroid belt.}
	\label{fig:edgeon}
\end{figure}

\begin{figure}
	\begin{center}                                                         
    	\includegraphics[width=1\columnwidth]{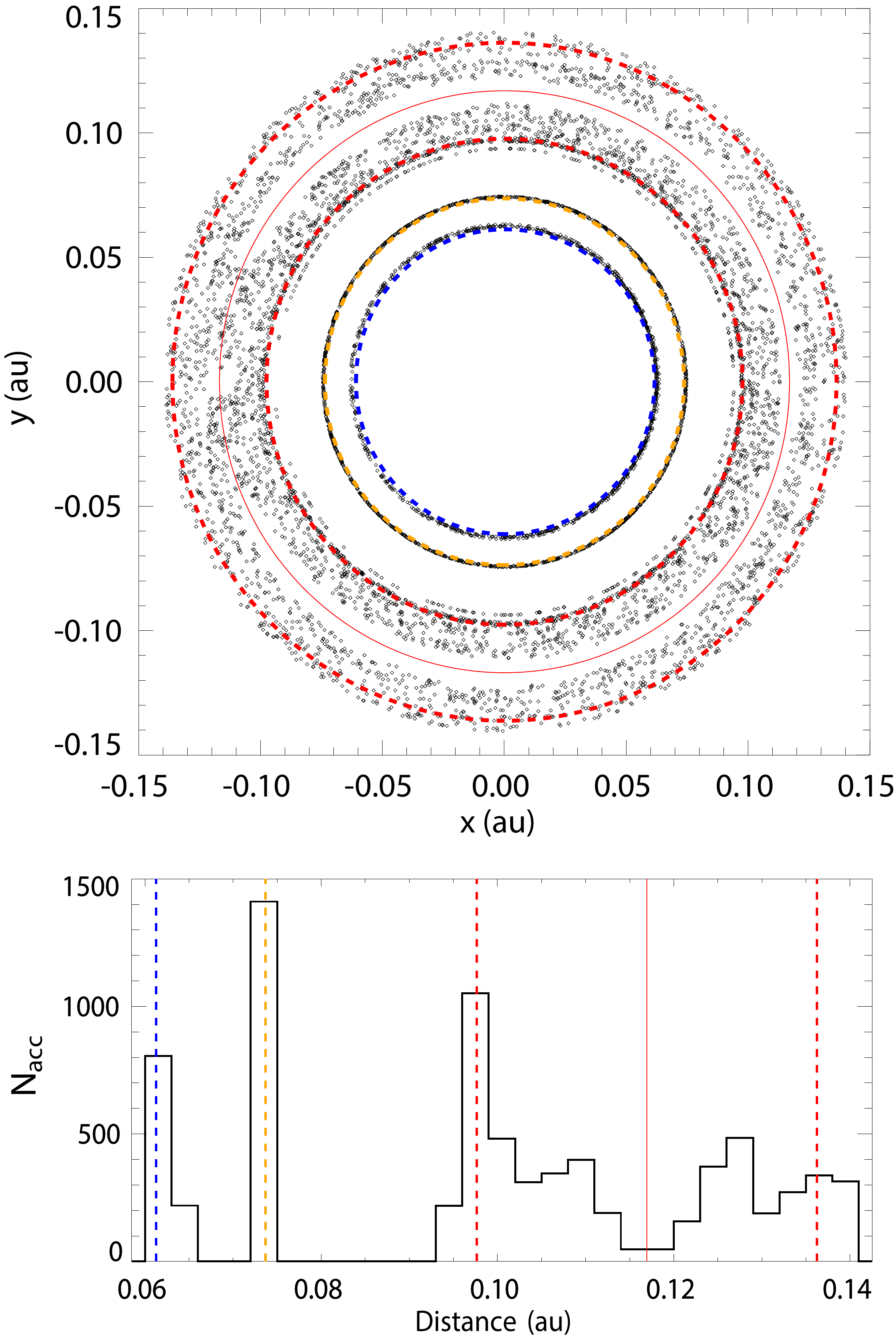}
    \end{center}
	\caption{Upper panel shows the initial distribution of accreted asteroids (indicated with black dots) in the model presented in Fig.\,\ref{fig:edgeon}. Red solid and red dashed lines show the orbit and the chaotic zone boundaries of planet\,i, respectively. Lower panel shows the histogram of the $N_\mathrm{acc}$ versus the initial distance of the asteroids. Majority of accreted asteroids are stem from the planet\,i's chaotic zone. However, a non-negligible amount of asteroids initially orbit at 1:2 MMR with planet\,i and the inner edge of the asteroid belt are also accreted in this particular model.}
	\label{fig:histo}
\end{figure}

To summary, our rersults show that the number of asteroids accreted by known TRAPPIST-1 planets decreases with the orbital distance and increases with the mass of the perturber. and increases with the orbital distance of the accreting planets. In order to understand the accretion phenomenons found in different models, we have to investigate the origin (initial orbital distance) of asteroids accreted by the planets. In the following, we focus on a particular model in which a 15\,M$_{\oplus}$ planet\,i orbits at 0.117\,au. Fig.\,\ref{fig:edgeon} shows edge-on view of the system emphasizing the region where the accreted asteroids originate from. Fig.\,\ref{fig:histo} upper panel shows the initial position of accreted asteroids, while the lower panel shows the histogram of the total number of accretion measured on planet\,b--h ($N_\mathrm{acc, tot}$) as a function of initial orbital distance of asteroids. 

As one can see, the majority of accreted asteroids initially orbit in the chaotic zone of planet\,i. Moreover, the number of accreted asteroids originated from the inner chaotic zone (inside the planetary orbit) is higher than that of the outer chaotic zone (outside the planetary orbit). However, Fig.\,\ref{fig:histo} reveals that some of the asteroids close to 1:1 MMR with planet\,i do not accreted by the known planets (see the horseshoe-like empty region on the \emph{upper panel} or the depression of $N_{\mathrm{acc, tot}}$ on the \emph{lower panel} formed at the orbital distance of planet\,i). See details later.

\begin{figure}
	\begin{center}                                                   
    	\includegraphics[width=1\columnwidth]{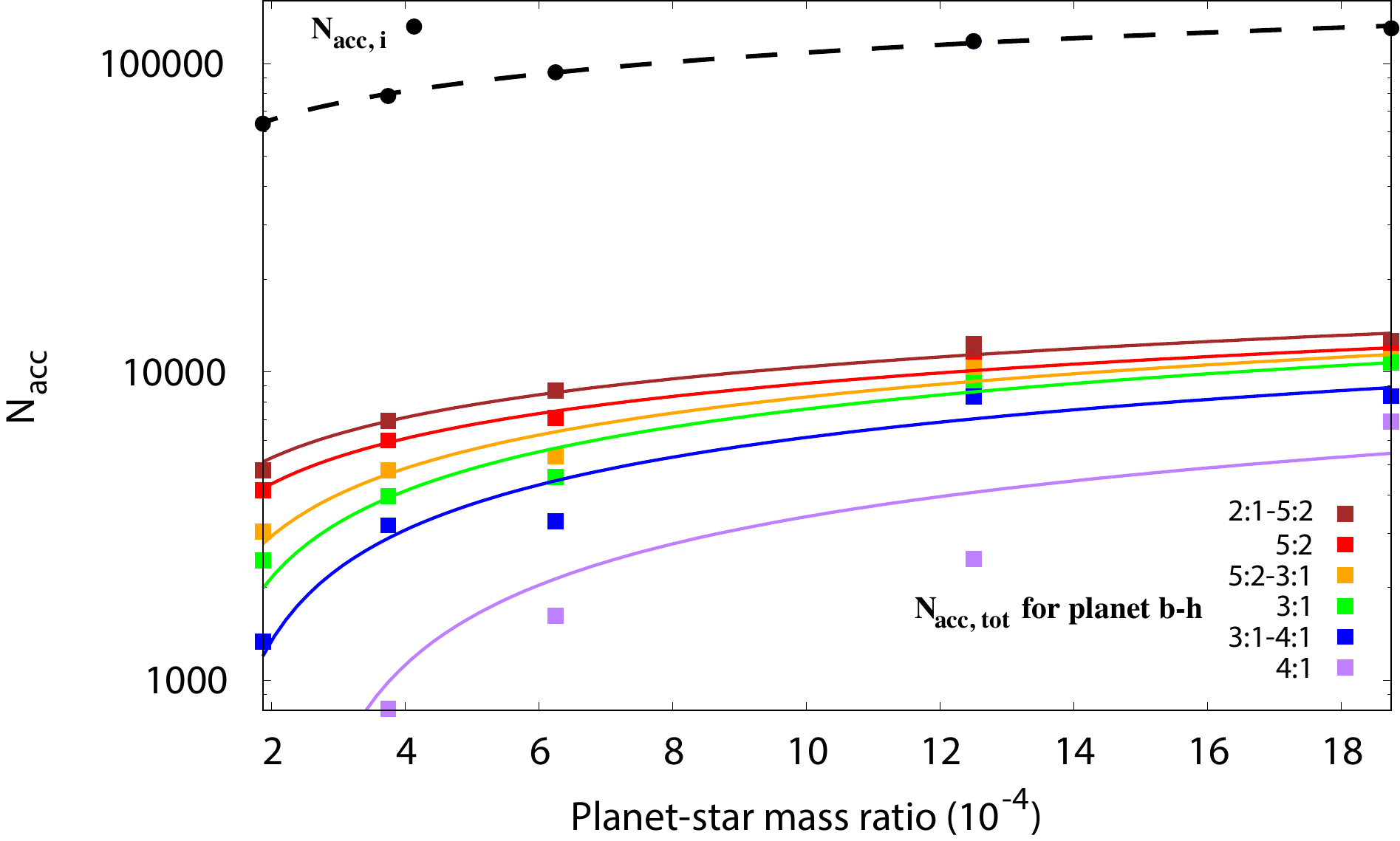}
	\end{center}
	\caption{Circles show $N_{\mathrm{acc, i}}$ as a function of planet\,i--star mass ratio. Squares indicate $N_{\mathrm{acc, tot}}$ on planet\,b--h as a function of planet\,i--star mass ratio with best-fitting curves. Colours represent different orbital distances of planet\,i.}
	\label{fig:total_massacc}
\end{figure}

It is plausible to assume that $N_{\mathrm{acc, tot}}$ is proportional to the number of asteroids originally orbit in planet\,i's chaotic zone, i.e. to the size of the chaotic zone. Fig.\,\ref{fig:total_massacc} shows the total number of accreted asteroids originated in the chaotic zone of planet\,i as a function of the mass ratio of planet\,i to TRAPPIST-1 ($q_{\mathrm{i}}$) assuming six different orbital distances (indicated in the key) of planet\,i. As one can see, the higher is the $m_{\mathrm{pl, i}}$, the larger is the total number of asteroids accreted by planet\,b--h for a given $a_{\mathrm{pl, i}}$. According to \citet{Wisdom1980} the chaotic zone size increases with $q_i^{2/7}$. Based on the refined fit to the chaotic zone size presented by \citet{Regalyetal2018}, we found that $N_{\mathrm{acc, tot}}\sim m_{\mathrm{pl}}^{0.36}$ reasonably fits the total accretion numbers to the known planets indicated by solid lines on Fig.\,\ref{fig:total_massacc}.

Although the size of the chaotic zone is linearly proportional to the orbital distance of planet\,i (see, e.g., \citealp{Wisdom1980}), we found that $N_{\mathrm{acc, tot}}$ decreases with $a_{\mathrm{pl, i}}$ (see Fig.\,\ref{fig:total_massacc}). This can be explained by the following train of thought. Initially, the number of the asteroids decreases with the distance as a function of $R^{-1}$ by assuming that the initial surface number density resembles to that of in a young asteroid belt \citep{Suetal2005}. As a result, the number of asteroids perturbed from the chaotic zone can be characterised by a constant value independent of $a_{\mathrm{pl, i}}$. However, this is note the case in our model. Let us assume that the eccentricity distribution of asteroids caused by the perturbation of planet\,i is independent of $a_{\mathrm{pl, i}}$. In this case, asteroids that can reach the inner system should have higher and higher eccentricities as $a_{\mathrm{pl, i}}$ increases. Since the number of asteroids that have sufficient eccentricity to enter to the inner system decreases with $a_{\mathrm{pl, i}}$, $N_{\mathrm{acc, tot}}$ decreases with $a_{\mathrm{pl, i}}$.

By analysing Fig.\,\ref{fig:histo} one can find that a non-negligible amount of asteroids initially orbits inside the planet\,i's chaotic zone do not collide to planet\,b--h. These asteroids either form stable islands incorporating the L4, L3, and L5 Lagrangian points or accreted by planet\,i itself. With increasing mass of planet\,i the number of asteroids captured in 1:1 MMR decreases: we found that asteroids initially orbit in L3 region are eventually accreted by planet\,b--h, i.e. they do not form stable island, if $m_{\mathrm{pl, i}}>15$\,M$_{\oplus}$. With regards the asteroids accreted by planet\,i, we found that $N_{\mathrm{acc, i}}$ as a function of $q_{\mathrm{i}}$ can be fitted by the size of planet\,i's Hill sphere, which is proportional to $q_\mathrm{i}^{1/3}$ (see dashed curve on Fig.\,\ref{fig:total_massacc}). This means that asteroids initially orbit in the Hill sphere of planet\,i are accreted by planet\,i. Since the size of the Hill sphere is proportional to $a_{\mathrm{pl, i}}$, while the number density distribution of the asteroids in our models is inversely proportional to $a_{\mathrm{pl, i}}$, the same $N_{\mathrm{acc, i}}(q_\mathrm{i})$ function can be given independent of $a_{\mathrm{pl, i}}$, see dashed curve in Fig.\,\ref{fig:total_massacc}. 

Note that planet\,i's chaotic zone is not the only source of asteroids accreted by planet\,b--h in the model shown in Fig.\,\ref{fig:histo}. A non-negligible amount (about 16\%) of accreted asteroids originated from 1:2 MMR with planet\,i shown by Fig.\,\ref{fig:histo} dashed orange line. The contribution of these sources to the accretion are tend to be negligible in models where planet\,i has a lower mass ($m_{\mathrm{pl, i}}<15$\,M$_{\oplus}$) and a larger orbital distance ($a_{\mathrm {pl, i}}>0.13$\,au). This can be explained by the fact that in these cases the excited eccentricities of asteroids are not high enough to enter the inner system. A known example of this phenomenon is the formation of 1:2 Kirkwood gap of the main asteroid belt in the Solar System: due to the perturbation of Jupiter asteroids can collide to planets, i.e. removed from the main asteroid belt \citep{MoonsMorbidelli1995,Tsiganisetal2002}.

Another source of the accrected asteroids should also be noted, namely the inner edge of the belt shown by dashed blue line in Fig.\,\ref{fig:histo}. The inner edge of the belt is set by the outer edge of the chaotic zone of planet\,h's assumed to be on nearly circular orbit ($e_{\mathrm{pl, h}}<0.05$). Due to a near-orbiting high mass perturber the orbit of planet\,h eccentricity increases slightly. As the size of the chaotic zone increases with $e_{\mathrm{pl, h}}$ the chaotic zone of planet\,h can enter to the asteroid belt. As a result, planet\,h can scatter a non-negligible amount of asteroids onto the known planets. This effect increases with the mass of planet\,i and decreases with $a_{\mathrm{pl, i}}$.

\subsection{Accretion to individual planets}

The number of the asteroids accreted by the known \mbox{TRAPPIST-1} planets increases with the orbital distance of the accreting planets (Fig.\,\ref{fig:totalacc}). As a result, most of the asteroids are accreted by planet\,h. Why can the inner planets accrete fewer number of asteroids than the outer ones? This is caused by two mutual effects. 

\begin{figure}
	\begin{center}                                                   
    	\includegraphics[width=1\columnwidth]{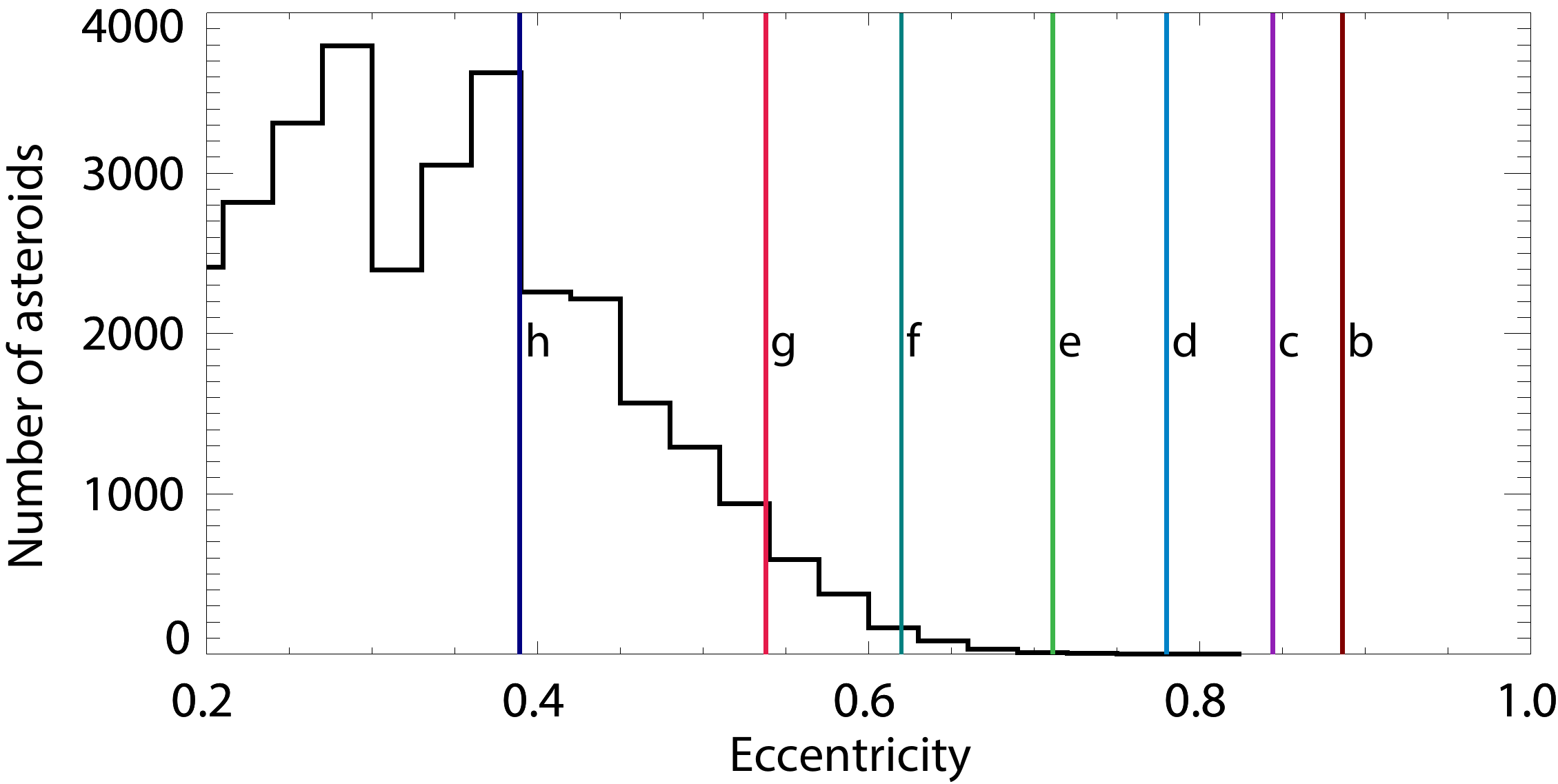}
	\end{center}
	\caption{$N_\mathrm{acc}$ as a function of the orbital eccentricity of asteroids after $5\times10^3$ orbit of planet\,b. Vertical lines show the critical eccentricities required for the asteroids to collide with a given planet of TRAPPIST-1 indicated with different colours.}
	\label{fig:histo_ecc}
\end{figure}

First, as we have shown the majority of asteroids accreted by planet\,b--h are originated from a given distance range, i.e. from the chaotic zone of the perturber. Therefore, in order to reach the inner planets asteroids have to be excited above critical eccentricities shown by coloured vertical lines in Fig.\,\ref{fig:histo_ecc}. Since the number of asteroids decreases towards higher eccentricities (see black line in Fig.\,\ref{fig:histo_ecc}), the number of available asteroids for accretion decreases with decreasing orbital distance of the accreting planets.

Second, every planet acts as a filter for the asteroid inflow flux. As a result, the farther the planet orbits in the TRAPPIST-1 system, the higher is the chance of removing asteroids from the system. To test this filter effect, we ran an additional model in which planet\,h was excluded, i.e. only six of the known planets and planet\,i were included. We found that the largest $N_{\mathrm{acc}}$ can be measured on planet\,g, i.e., the outermost planet of this particular model.

\subsection{Estimation of transported water mass}

Based on atmospheric transmission spectroscopic measurements \citep{deWitetal2018} the atmosphere of TRAPPIST-1 planets (e, f, g, and h) can be rich in water. Density estimations of planets also suggest the presence of water on planet\,e, f, and g. Since they orbit in the habitable zone water oceans can be in liquid state. However, the estimation of planet radius (required for density determination) based on transit depth measurements \citep{Gillonetal2017} might be uncertain due to inhomogeneous distribution of stellar spots on TRAPPIST-1 photosphere \citep{Rackhametal2018}. As a result, the density of the planets is larger by about 3 percent than estimated by \cite{Gillonetal2017}, which corresponds to a lower water content than it was previously expected.  

In order to determine the mass of the asteroids accreted by a given planet, first the total mass of the initial asteroid belt should be given. Our estimations based on the minimum mass nebula for M dwarfs calculations of \cite{Gaidos2017} and the surface density estimation of \cite{Kraletal2018} for a planetesimal belt in the TRAPPIST-1 system. Our calculations show that the mass of the asteroid ring lying between 0.061\,au and 0.256\,au is about $1$\,M$_{\oplus}$. Therefore, the mass of an individual asteroid is $2\times 10^{-6}$\,M$_{\oplus}$ resembling to that $5\times10^5$ asteroids are included in our models. Assuming that these asteroids are similar to centaurs in our Solar System, whose water fraction is about 50 percent \citep{Cruikshanketal1998,Schwarzetal2017}, any given asteroids can transport $10^{-6}$\,M$_{\oplus}$ water. 

\begin{figure}
	\begin{center}                                                     
    	\includegraphics[width=1\columnwidth]{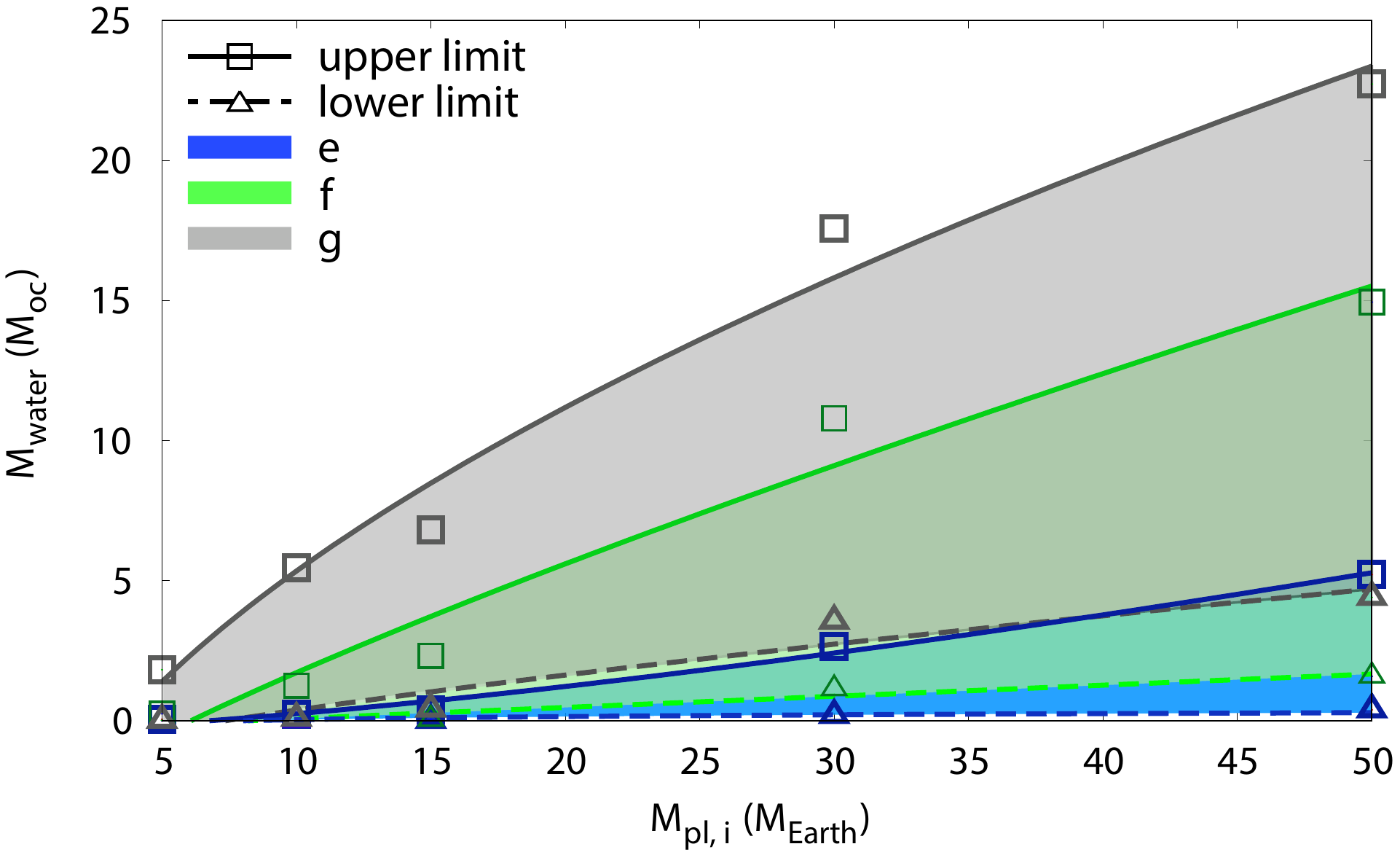}
	\end{center}
	\caption{Accreted water mass (measured in Earth ocean mass units) as a function of the mass of planet\,i. Squares and triangles indicate the minima (for the largest $a_{\mathrm{pl, i}}$) and maxima (for the smallest $a_{\mathrm{pl, i}}$) of the water mass values, respectively. Shaded regions with different colours refer to planet\,e (blue), f (green), and g (grey).}
	\label{fig:wateracc}
\end{figure}

Fig.\,\ref{fig:wateracc} shows the estimated water mass transported to the habitable planets measured in units of Earth ocean mass (M$_{\mathrm{oc}}=2.34\times10^{-4}\,\mathrm{M}_{\oplus}$) as a function of $m_{\mathrm{pl, i}}$. Solid and dashed lines indicate the fitted upper and lower limit of the water mass accreted by planet\,e, f, and g, respectively. The maximum and minimum limits correspond to models where the given mass planet\,i orbits at the smallest and largest distances, respectively. Based on $N_{\mathrm{acc}}$ values the maximum water masses are the followings in the order of ocean mass: planet\,g: 22.73\,M$_{\mathrm{oc}}$; planet\,f: 14.94\,M$_{\mathrm{oc}}$; planet\,e: 5.22\,M$_{\mathrm{oc}}$. The minimum values are the followings: planet\,g: 0.01\,M$_{\mathrm{oc}}$; both of planet\,f and e: 0\,M$_{\mathrm{oc}}$. The shaded regions of Fig.\,\ref{fig:wateracc} show the masses between the extreme values.

Emphasize that the estimated water mass values are uncertain as our calculation based on the initial mass of the putative asteroid belt. Namely, the amount of water transported are scaled with the mass of the belt. However, as it is shown in Fig.\,\ref{fig:wateracc}, the amount of water accreted by TRAPPIST-1 planets increases with their orbital distance in all models independent of $m_\mathrm{pl,i}$ and $a_\mathrm{pl,i}$.

As we shown, among the habitable planets, TRAPPIST-1g can accrete the highest amount of water. Based on the analysis of transit and TTV signals, the mass and the mean density of planet\,g are found to be slightly higher (1.148\,M$_{\oplus}$) and lower (0.759\,$\rho_{\oplus}$) than that of Earth according to \citet{Grimmetal2018}. These results suggest that planet\,g might have a higher mass water reservoir than Earth has, which is in agreement with our estimations. Planet\,f has a lower mass (0.932\,M$_{\oplus}$) and also lower density (0.816\,$\rho_{\oplus}$) than that of Earth suggesting that it has larger mass oceans than Earth has. Although, planet\,e's mass is 0.772 times lower, while its density is 1.024 times higher than that of Earth, the water content of this planet could be the lowest among the habitable TRAPPIST-1 planets.

According to \cite{Grimmetal2018} the densities of the habitable planets (planet\, e, f, and g) and planet\,h decrease with the semi-major axis of the planetary orbits, which is consistent with our results of increasing amount of accreted water with the orbital distance. According to our simulations, the planetary densities can be reduced by about 2.5 percent on average due to an LHB-like water delivery assuming the strongest asteroid bombardment model (see Fig.\,\ref{fig:wateracc}). The maximum density reduction can occur for planet\,h, which is found to be about 12.44 percent in our model. However, the spread in the density values of TRAPPIST-1 planets is observed to be 17\,percent \citep{Grimmetal2018}. Thus, to explain the density variations within the confines of our model the mass of asteroid belt should be at least two times larger. Moreover, it is known that the density of the innermost three planets do not follow the above trend. Therefore, we conclude that the observed trend in the densities is presumably not caused by a hypothetical LHB-like event.

\section{Conclusion}

In this paper we investigated a putative late heavy bombardment like event in TRAPPIST-1 system, which can transport a significant amount of water to the habitable planets discovered in the system. We assume that the source of the asteroids is a belt of water-rich small bodies beyond the snow line (0.049\,au). In our hypothetical model, a putative planet having relatively high mass compared to the known planets, perturbs the orbit of asteroids. First, we investigated the stability of this hypothetical system by means of high-precision N-body (N=9) simulations done with IAS15 \citep{ReinTamayo2016}. Stability investigations span $3.85\times10^{7}$ orbits of planet\,h corresponding to $4.8\times10^8$ orbits of planet\,b. Following the stability analysis, we modelled the TRAPPIST-1 system extended with a putative planet and an asteroid belt containing half a million massless bodies by means of 9 plus 1 body problem (HIPERION). We run simulations assuming dynamically cold and hot asteroid belts. In the hot scenario asteroids have initial eccentricity and inclination. We investigated the effect of the mass (5--50\,M$_{\oplus}$), orbital distance (0.078--0.197\,au), and orbital inclination (0--5\degree) of the putative planet (planet\,i) on the asteroid impact flux measured on the known TRAPPIST-1 planets. 
Our main results concerning the stability of the extended systems are the followings:
\begin{enumerate}
    \item Planet\,i orbiting beyond 0.162\,au (4:1 MMR with planet\,h)  only slightly influences the orbital elements of TRAPPSIT-1 planets, i.e these systems are stable. 
    \item However, if planet\,i is on a closer orbit, the extended systems are stable only for the following cases: $a_{\mathrm{pl, i}}=0.086$\,au for $m_{\mathrm{pl, i}}\leq10$\,M$_\oplus$, $a_{\mathrm{pl, i}}=0.102$\,au for $m_{\mathrm{pl, i}}\leq15$\,M$_\oplus$, $a_{\mathrm{pl, i}}=0.117$\,au for $m_{\mathrm{pl, i}}\leq30$\,M$_\oplus$.
\end{enumerate}
	
Our main results on the asteroid impact flux to the known TRAPPIST-1 planets are the followings:
\begin{enumerate}
    \item The number of asteroids accreted by TRAPPIST-1 known planets is nearly independent of the initial eccentricity and inclination of asteroids (i.e. dynamical temperature) being in the investigated range. The time-scale of LHB-like event found to be 1-2 orders of magnitude longer in the dynamically hot than in the dynamically cold asteroid belt models.
	\item Initially, the accretion rates suddenly increase then saturate. The saturation time-scale of accretion rates, i.e. the longevity of the LHB-like event, decreases with the orbital distance of accreting planet and increases with the orbital distance and decreasing mass of planet\,i.
	\item The total number of asteroids accreted by a given planet increases with the mass and decreases with the orbital distance of planet\,i.
    \item A definite relation exists between the chaotic zone size of planet\,i and the total number of accreted asteroids by the known planets. This is because of the fact that the majority of asteroids accreted by the known planets are originated from the chaotic zone of planet\,i.
    \item However, in some cases ($m_{\mathrm{pl, i}}>15$\,M$_{\oplus}$ and $a_{\mathrm {pl, i}}<0.13$\,au) a non-negligible amount of accreted asteroids are originated from 1:2 MMR with planet\,h. This is because of the fact that the excited eccentricity of asteroids originally orbiting at 1:2 MMR are sufficiently high to reach the inner system. 
    \item The number of asteroids accreted by a given planet is proportional to the orbital distance of that planet. As a result, the farther the planet orbits TRAPPIST-1, the larger is the asteroid impact flux.   
\end{enumerate}

Here we have to mention some caveats of our hypothetical LHB-like event. First, asteroids are represented by massless particles. This approximation excludes the possibility of planetesimal-driven asteroid migration, which could increase the number of accreted asteroids in the inner system \citep{Bonsoretal2014,RaymondBonsor2014}. The gravitational effect of the asteroid belt assumed to have about 1\,M$_\oplus$ was neglected. Since the gravitational perturbation of the belt itself can influence both planetary and asteroid orbits it is worth investigating that effect in a future study.

Second, our estimation for the water mass transported is highly uncertain as the mass of the asteroid belt is based on a simple extrapolation of the minimum mass solar nebula model to M dwarf stars. Emphasize that our estimation for the water content based on dynamically cold models where the orbits of asteroids and TRAPPIST-1 planets are assumed to be coplanar. Therefore, our results can only be interpreted as an upper limit. Our models do not take into account the water loss due to the collisional heat caused by the asteroid impacts (see, e.g., \citealp{Kraletal2018}). Moreover, the asteroid belt is assumed to be primordial, which does not go under any mass depletion processes, i.e. formation of planets.

Finally, our model is oversimplified. We do not investigate the role of planetary migration. Planets can form in situ if there are enough dust in the protoplanetary disc in the vicinity of their current orbit. A possible solution to this problem is the inward migration of pebbles, which later become the building blocks of planets  \citep{HansenMurray2012}. However, the formation of the resonant chain between the planetary periods is hardly conceivable without a resonant capturing via inward migration of planets \citep{Lugeretal2017}. Another possible solution is that TRAPPIST-1 planets formed beyond the snow line then migrated inwards. \citet{Ormeletal2017} have shown that the resonant chain of planets can be formed in such process. Planet formation began beyond the snow line where the water condensates from the gas phase of the protoplanetary disc \citep{Unterbornetal2018}. Although planets could also collect water-rich planetesimals during their migration, the preservation efficiency of volatile content during intense collisions are unknown. In this scenario the migration of planet\,i can cause higher asteroid impact flux than our estimations as planet\,i's chaotic zone can be continuously replenished by subsequent asteroids during the migration across the asteroid belt. Note that the asteroid belt moves outwards \citep{Izidoroetal2014} during the epoch of inward planetary migration. Therefore, the inner edge of the asteroid belt might lie farther out than it is assumed in our models. However, this has only a minor effect on the accretion numbers as majority of the accreted asteroids originates from the chaotic zone of planet\,i not from the inner edge of the belt.

Based on our results, we conclude that the efficiency of water transport increases with the orbital distance of known TRAPPIST-1 planets. Therefore, the mass of water delivered by a putative LHB-like event increases with the orbital distance of planets. Although our findings suggest that the densities of planets decrease with their orbital distances, the amount of water accreted is negligible to the planetary masses. Thus, the observable trend in planetary densities could not be caused by a hypothetical LHB-like event. Important to note another problem with an LHB-like event is that some studies (e.g., \citealp{Gomesetal2005,OBrienetal2018}) suggest that only a small fragment (1--10\%) of the total mass of the oceans was delivered by asteroids in the case of Earth.

As we have shown, planet\,h collects the largest amount of water by an LHB-like event among TRAPPIST-1 planets. Although planet\,h orbits beyond the habitable zone, it may host life in a similar environment to Jupiter's moon Europa. The surface of Europa is covered by ice crust. Under the crust there can be found a subsurface ocean according to density measurements \citep{KuskovKronrod2005}. The subsurface water could be in liquid phase due to the high pressure of the crust and the tidal heating from the gravitational interactions between the moon and Jupiter. Under the ice sheet of planet\,h a water ocean might exist due to the tidal heating of TRAPPIST-1 planets being in a resonance chain \citep{Lugeretal2017}. 

According to our simulations the amount of water transported via an LHB-like event increases with the orbital distance of accreting planets. It means that asteroid fluxes measured on a system of planets in an LHB-like event depend on the orbital distances of planets. Thus, our conclusions are applicable to the Solar System to get more accurate estimations for the meteoritic impact numbers of rocky planets than the currently accepted crater density value \cite{Ivanov2001}.

\subsection*{Acknowledgement}

This work was supported by the Hungarian Grant K119993. We also acknowledge NIIF for awarding us access to computational resource based in Hungary at Debrecen. Furthermore, we are obliged for the offered Cloud service of CPU threads by Wigner Research Centre for Physics. ZD thanks to K. Vida, T. Kov\'acs and A. Kereszturi for helpful discussions. The authors also thank to the anonymous referee for thoughtful comments that helped to improve the quality of the paper.

\label{lastpage}


\begin{thebibliography}{3}

\bibitem[\protect\citeauthoryear{Andersen \& Korhonen}{2015}]{AndersenKorhonen2015} Andersen J.~M., Korhonen H., 2015, MNRAS, 448, 3053
\bibitem[\protect\citeauthoryear{Bolmont et al.}{2017}]{Bolmontetal2017} Bolmont E., Selsis F., Owen J.~E., Ribas I., Raymond S.~N., Leconte J., Gillon M., 2017, MNRAS, 464, 3728 
\bibitem[\protect\citeauthoryear{Bonsor, Augereau, \& Th{\'e}bault}{2012}]{Bonsoretal2012} Bonsor A., Augereau J.-C., Th{\'e}bault P., 2012, A\&A, 548, A104 
\bibitem[\protect\citeauthoryear{Bonsor et al.}{2013}]{Bonsoretal2013} Bonsor A., Kennedy G.~M., Crepp J.~R., Johnson J.~A., Wyatt M.~C., Sibthorpe B., Su K.~Y.~L., 2013, MNRAS, 431, 3025 
\bibitem[\protect\citeauthoryear{Bonsor et al.}{2014}]{Bonsoretal2014} Bonsor A., Raymond S.~N., Augereau J.-C., Ormel C.~W., 2014, MNRAS, 441, 2380 
\bibitem[\protect\citeauthoryear{Boss et al.}{2017}]{Bossetal2017} Boss A.~P., Weinberger A.~J., Keiser S.~A., Astraatmadja T.~L., Anglada-Escude G., Thompson I.~B., 2017, AJ, 154, 103 
\bibitem[\protect\citeauthoryear{Bottke et al.}{2012}]{Bottkeetal2012} Bottke W.~F., Vokrouhlick{\'y} D., Minton D., Nesvorn{\'y} D., Morbidelli A., Brasser R., Simonson B., Levison H.~F., 2012, Natur, 485, 78
\bibitem[\protect\citeauthoryear{Bourrier et al.}{2017}]{Bourrieretal2017} Bourrier V., et al., 2017, AJ, 154, 121
\bibitem[\protect\citeauthoryear{Cruikshank et al.}{1998}]{Cruikshanketal1998} Cruikshank D.~P., et al., 1998, Icar, 135, 389 
\bibitem[\protect\citeauthoryear{de Wit et al.}{2018}]{deWitetal2018} de Wit J., et al., 2018, NatAs, 2, 214
\bibitem[\protect\citeauthoryear{Gaidos}{2017}]{Gaidos2017} Gaidos E., 2017, MNRAS, 470, L1 
\bibitem[\protect\citeauthoryear{Gillon et al.}{2017}]{Gillonetal2017} Gillon M., et al., 2017, Natur, 542, 456 
\bibitem[\protect\citeauthoryear{Gomes et al.}{2005}]{Gomesetal2005} Gomes R., Levison H.~F., Tsiganis K., Morbidelli A., 2005, Natur, 435, 466 
\bibitem[\protect\citeauthoryear{Grimm et al.}{2018}]{Grimmetal2018} Grimm S.~L., et al., 2018, A\&A, 613, A68 
\bibitem[\protect\citeauthoryear{Hahn}{2003}]{Hahn2003} Hahn J.~M., 2003, ApJ, 595, 531
\bibitem[\protect\citeauthoryear{Hansen \& Murray}{2012}]{HansenMurray2012} Hansen B.~M.~S., Murray N., 2012, ApJ, 751, 158 
\bibitem[\protect\citeauthoryear{Ida \& Makino}{1992}]{IdaMakino1992} Ida S., Makino J., 1992, Icar, 96, 107 
\bibitem[\protect\citeauthoryear{Ivanov}{2001}]{Ivanov2001} Ivanov B.~A., 2001, SSRv, 96, 87
\bibitem[\protect\citeauthoryear{Izidoro, Morbidelli, \& Raymond}{2014}]{Izidoroetal2014} Izidoro A., Morbidelli A., Raymond S.~N., 2014, ApJ, 794, 11 
\bibitem[\protect\citeauthoryear{Izidoro et al.}{2017}]{Izidoroetal2017} Izidoro A., Ogihara M., Raymond S.~N., Morbidelli A., Pierens A., Bitsch B., Cossou C., Hersant F., 2017, MNRAS, 470, 1750 
\bibitem[\protect\citeauthoryear{Kral et al.}{2018}]{Kraletal2018} Kral Q., Wyatt M.~C., Triaud A.~H.~M.~J., Marino S., Th{\'e}bault P., Shorttle O., 2018, MNRAS, 479, 2649 
\bibitem[\protect\citeauthoryear{Kuskov \& Kronrod}{2005}]{KuskovKronrod2005} Kuskov O.~L., Kronrod V.~A., 2005, Icar, 177, 550 
\bibitem[\protect\citeauthoryear{Luger et al.}{2017}]{Lugeretal2017} Luger R., et al., 2017, NatAs, 1, 0129 
\bibitem[\protect\citeauthoryear{Marcy et al.}{2014}]{Marcyetal2014} Marcy G.~W., et al., 2014, ApJS, 210, 20 
\bibitem[\protect\citeauthoryear{Mocquet, Grasset, \& Sotin}{2014}]{Mocquetetal2014} Mocquet A., Grasset O., Sotin C., 2014, RSPTA, 372, 20130164
\bibitem[\protect\citeauthoryear{Moons \& Morbidelli}{1995}]{MoonsMorbidelli1995} Moons M., Morbidelli A., 1995, Icar, 114, 33 
\bibitem[\protect\citeauthoryear{Morbidelli et al.}{2000}]{Morbidellietal2000} Morbidelli A., Chambers J., Lunine J.~I., Petit J.~M., Robert F., Valsecchi G.~B., Cyr K.~E., 2000, M\&PS, 35, 1309 
\bibitem[\protect\citeauthoryear{O'Brien et al.}{2018}]{OBrienetal2018} O'Brien D.~P., Izidoro A., Jacobson S.~A., Raymond S.~N., Rubie D.~C., 2018, SSRv, 214, 47 
\bibitem[\protect\citeauthoryear{Ormel, Liu, \& Schoonenberg}{2017}]{Ormeletal2017} Ormel C.~W., Liu B., Schoonenberg D., 2017, A\&A, 604, A1 
\bibitem[\protect\citeauthoryear{Papaloizou, Szuszkiewicz, \& Terquem}{2018}]{Papaloizouetal2018} Papaloizou J.~C.~B., Szuszkiewicz E., Terquem C., 2018, MNRAS, 476, 5032 
\bibitem[\protect\citeauthoryear{Press \& Spergel}{1988}]{PressSpergel1988} Press W.~H., Spergel D.~N., 1988, ApJ, 325, 715 
\bibitem[\protect\citeauthoryear{Quarles et al.}{2017}]{Quarlesetal2017} Quarles B., Quintana E.~V., Lopez E., Schlieder J.~E., Barclay T., 2017, ApJ, 842, L5
\bibitem[\protect\citeauthoryear{Rackham, Apai, \& Giampapa}{2018}]{Rackhametal2018} Rackham B.~V., Apai D., Giampapa M.~S., 2018, ApJ, 853, 122
\bibitem[\protect\citeauthoryear{Raymond, Quinn, \& Lunine}{2007}]{Raymondetal2007} Raymond S.~N., Quinn T., Lunine J.~I., 2007, AsBio, 7, 66 
\bibitem[\protect\citeauthoryear{Raymond, Armitage, \& Gorelick}{2010}]{Raymondetal2010} Raymond S.~N., Armitage P.~J., Gorelick N., 2010, ApJ, 711, 772
\bibitem[\protect\citeauthoryear{Raymond et al.}{2011}]{Raymondetal2011} Raymond S.~N., et al., 2011, A\&A, 530, A62 
\bibitem[\protect\citeauthoryear{Raymond \& Bonsor}{2014}]{RaymondBonsor2014} Raymond S.~N., Bonsor A., 2014, MNRAS, 442, L18 
\bibitem[\protect\citeauthoryear{Raymond \& Izidoro}{2017}]{RaymondIzidoro2017} Raymond S.~N., Izidoro A., 2017, Icar, 297, 134 
\bibitem[\protect\citeauthoryear{Reg{\'a}ly et al.}{2018}]{Regalyetal2018} Reg{\'a}ly Z., Dencs Z., Mo{\'o}r A., Kov{\'a}cs T., 2018, MNRAS, 473, 3547 
\bibitem[\protect\citeauthoryear{Rein \& Spiegel}{2015}]{ReinSpiegel2015} Rein H., Spiegel D.~S., 2015, MNRAS, 446, 1424
\bibitem[\protect\citeauthoryear{Rein \& Tamayo}{2016}]{ReinTamayo2016} Rein H., Tamayo D., 2016, MNRAS, 459, 2275 
\bibitem[\protect\citeauthoryear{Reynolds, McKay, \& Kasting}{1987}]{Reynoldsetal1987} Reynolds R.~T., McKay C.~P., Kasting J.~F., 1987, AdSpR, 7, 125 
\bibitem[\protect\citeauthoryear{Schwarz, Bazs{\'o}, \& Loibnegger}{2017}]{Schwarzetal2017} Schwarz R., Bazs{\'o} {\'A}., Loibnegger B., 2017, hccm.conf, \#1 
\bibitem[\protect\citeauthoryear{Shields, Ballard, \& Johnson}{2016}]{Shieldsetal2016} Shields A.~L., Ballard S., Johnson J.~A., 2016, PhR, 663, 1
\bibitem[\protect\citeauthoryear{Su et al.}{2005}]{Suetal2005} Su K.~Y.~L., et al., 2005, ApJ, 628, 487 
\bibitem[\protect\citeauthoryear{Tamayo et al.}{2017}]{Tamayoetal2017} Tamayo D., Rein H., Petrovich C., Murray N., 2017, ApJ, 840, L19
\bibitem[\protect\citeauthoryear{Tanner et al.}{2012}]{Tanneretal2012} Tanner A., White R., Bailey J., Blake C., Blake G., Cruz K., Burgasser A.~J., Kraus A., 2012, ApJS, 203, 10 
\bibitem[\protect\citeauthoryear{Thommes et al.}{2008}]{Thommesetal2008} Thommes E.~W., Bryden G., Wu Y., Rasio F.~A., 2008, ApJ, 675, 1538 
\bibitem[\protect\citeauthoryear{Tsiganis, Varvoglis, \& Hadjidemetriou}{2002}]{Tsiganisetal2002} Tsiganis K., Varvoglis H., Hadjidemetriou J.~D., 2002, Icar, 159, 284 
\bibitem[\protect\citeauthoryear{Unterborn et al.}{2018}]{Unterbornetal2018} Unterborn C.~T., Desch S.~J., Hinkel N.~R., Lorenzo A., 2018, NatAs, 2, 297 
\bibitem[\protect\citeauthoryear{Vida et al.}{2017}]{Vidaetal2017} Vida K., K{\H o}v{\'a}ri Z., P{\'a}l A., Ol{\'a}h K., Kriskovics L., 2017, ApJ, 841, 124 
\bibitem[\protect\citeauthoryear{Walsh et al.}{2011}]{Walshetal2011} Walsh K.~J., Morbidelli A., Raymond S.~N., O'Brien D.~P., Mandell A.~M., 2011, Natur, 475, 206 
\bibitem[\protect\citeauthoryear{Wang et al.}{2017}]{Wangetal2017} Wang S., Wu D.-H., Barclay T., Laughlin G.~P., 2017, arXiv, arXiv:1704.04290 
\bibitem[\protect\citeauthoryear{Williams \& Cieza}{2011}]{WilliamsCieza2011} Williams J.~P., Cieza L.~A., 2011, ARA\&A, 49, 67
\bibitem[\protect\citeauthoryear{Wisdom}{1980}]{Wisdom1980} Wisdom J., 1980, AJ, 85, 1122 

\end{thebibliography}
\end{document}